\documentclass[%
 reprint,
showpacs,preprintnumbers,
 amsmath,amssymb,
 aps,
]{revtex4-1}

\usepackage{graphicx}
\usepackage{dcolumn}
\usepackage{bm}
\usepackage{xcolor}

\usepackage{dsfont} 

\newcommand{\mathH}{\hat{\mathcal{H}}}

\begin{document}


\title{Solvable Loop Quantum Cosmology: domain of the volume observable and semiclassical states}

\author{Mercedes Mart\'{i}n-Benito}%
 \email{m.martin.benito@ucm.es}
 \affiliation{%
 Departamento de F\'isica Te\'orica,  Universidad Complutense de Madrid,  Parque de Ciencias 1,   28040 Madrid, Spain
}%

\author{Rita B. Neves}%
 \email{rbneves@fc.ul.pt}
\affiliation{%
 Instituto  de  Astrof\'{i}sica  e  Ci\^{e}ncias  do  Espa\c{c}o,  Faculdade  de  Ci\^{e}ncias  da Universidade  de  Lisboa,  Campo  Grande,  PT1749-016  Lisboa,  Portugal
}%

\date{\today}

\begin{abstract}
The dynamics of a flat Friedmann-Lema\^{i}tre-Robertson-Walker model minimally coupled to a massless scalar field has been intensively studied in the context of Loop Quantum Cosmology. This model admits an appropriate solvable representation, named sLQC. The form of the domain of the volume, the main observable to track the quantum evolution, is not straightforward in this solvable representation, and its explicit construction has been overlooked so far. In this work we find the explicit form of physical states belonging to the domain of the volume in  sLQC. Specifically, given a physical state in the $v$-representation where the volume acts diagonally, we derive its form in the representation employed in sLQC, making explicit the connection between both representations at the physical level. To this end, we resort to the Wheeler-De Witt (WDW) approach, which shares the physical Hilbert space with sLQC when cast in an analog solvable representation, while being analytically solvable as well in the $v$-representation. Then the domain of the volume for the WDW approach provides that for sLQC. Furthermore, we address the question of semiclassicality in sLQC.
\end{abstract}

\maketitle


\section{Introduction}

The field of quantum cosmology has experienced in recent years a boost in activity, partly motivated by the fact that the physics of the early Universe could reveal effects coming from the quantum nature of spacetime, and thus provide a potential observational window to quantum gravity. Although several different approaches are being explored, one of the most promising ones is Loop Quantum Cosmology (LQC) \cite{LQCreview_Bojowald2005, LQCreview_Banerjee2012, LQCreview_Agullo2016, LQCreview_Ashtekar2011}. Besides leading to interesting predictions, like the resolution of the big-bang singularity by means of a quantum bounce, it is a well motivated approach guided by a full proposal for non-perturbative quantum gravity: Loop Quantum Gravity (LQG) \cite{LQGreview_Ashtekar2004, Rovelli_QG,Thiemann_notes}.

The first cosmology quantized in the context of LQC was a flat Friedmann-Lema\^{i}tre-Robertson-Walker (FLRW) model minimally coupled to a massless scalar field $\phi$ for the matter sector \cite{APS,APS_improvedDyn,APS_improvedDyn_PRD}. This is the simplest cosmological model, consisting of two degrees of freedom in configuration space (one for the geometry and one for the matter) subject to a single constraint, namely the zero mode of the Hamiltonian constraint of General Relativity (GR) particularized for this model. The geometry degree of freedom was quantized employing a non-standard polymeric representation \cite{Ashtekar_Bojowald_Lewandowski}, following the techniques of LQG, which leads to a discretization of the geometry. Given this canonical approach, one finds the physical Hilbert space as the space of solutions of the quantum Hamiltonian constraint endowed with an appropriate inner product. The model was numerically solved for certain states in \cite{APS_improvedDyn,APS_improvedDyn_PRD}  following the so-called \textit{improved dynamics} procedure. In the chosen representation, the volume acts by multiplication, and we will call it $v$-representation.
Interpreting $\phi$ as the internal time variable, the work of \cite{APS_improvedDyn,APS_improvedDyn_PRD} shows that the quantum evolution (with respect to the relational clock $\phi$) of the expectation values of the volume and energy density observables on physical semiclassical states never diverge. 
Instead, when the energy density approaches a critical value, of the order of the Planck scale, the quantum effects of the geometry render gravity repulsive and the energy density and the volume observables undergo a bounce. The energy density reaches a finite maximum and the volume a non-zero minimum, connecting a contracting epoch of the Universe with an expanding one. Furthermore, this procedure proved to be robust, as it was also successfully applied to models with non-zero cosmological constant \cite{APS_improvedDyn, Bentivegna2008, Kaminski2009, Pawlowski2012}, spatially compact models \cite{K=1Cosmologies, closedFRWcosmologies}, and Bianchi models \cite{Merce_bianchiI, Ashtekar_bianchiI, Ashtekar_bianchiII, WilsonEwing_bianchiIX}.

Then, the concern fell on whether the bounce was particular to the semiclassical states analyzed numerically in \cite{APS_improvedDyn,APS_improvedDyn_PRD}  or a general feature of the approach. This question was answered in \cite{ACS}, for the simplest case of the flat FLRW model with a massless scalar field. This work introduced a convenient change of representation that casts the Hamiltonian constraint into a Klein-Gordon equation in $1+1$ dimensions,  in which again one interprets the scalar field as the time variable.
This way, the model turns out to be exactly solvable in terms of left and right moving modes, and this formulation was named solvable LQC (sLQC).  Remarkably, the bounce is proven to occur for a generic physical state, as a consequence of the discreteness of the geometry. 

However, there is still an open question that has been overlooked hitherto, and that we address in this work. Namely, the question of which physical states in the solvable representation of LQC belong to the domain of the volume, which is the main observable under consideration to track the dynamics of the system. This question is non-trivial since the volume is an unbounded (essentially self-adjoint) operator that in sLQC no longer acts by multiplication. Because the sLQC prescription was introduced after the bounce had already been found to occur numerically, it was not necessary to construct explicitly states in the domain of the volume. As a consequence, the fact that it is actually non-straightforward has gone unnoticed. 

Our interest of providing the explicit form of the states forming the domain of the volume goes beyond completeness of the formulation of sLQC. Having that knowledge proves to be useful when trying to integrate the dynamics of other non-solvable models. Indeed, reference  \cite{Merce_MSeq} proposes an approximation method designed for the flat FLRW model coupled to a scalar field with a potential, such as an inflaton, which generically does not admit an analytical solution. This approach is based on perturbation theory for systems with time-dependent Hamiltonians, and introduces an interaction picture that relies on the integration of the ``free dynamics'' of the solvable model (scalar field without potential). In order to apply the method, one needs to be able to provide explicit physical states of sLQC that belong to the domain of the volume, and moreover that are semiclassical, to ensure that far from the high curvature regimes the physics of the model resembles the classical one predicted by GR. 

Even though  \cite{ACS} gives the expression of the expectation value of the volume on physical states in sLQC, it is obvious that it is not well-defined for general physical states. Furthermore, as far as we know the question of semiclassicality in this formulation has never been addressed either. Answering these questions requires making the connection between the physical Hilbert space of sLQC and the physical Hilbert space of the original $v$-representation of LQC employed in \cite{APS_improvedDyn,APS_improvedDyn_PRD}. In  \cite{ACS} this connection is made at the kinematical level, but not at the physical one.
In this work, we precisely provide this connection. Starting with a physical profile in the original $v$-representation of LQC, we manage to work out the expression of that physical state in the Klein-Gordon formulation of sLQC, which allows us, in particular, to write semiclassical states in both formulations.

We find that this connection is more easily obtained in the context of the Wheeler-De Witt (WDW) approach. In this approach, instead of the polymeric quantization of LQC, a standard Schr\"{o}dinger-like representation is adopted. It also allows for a Klein-Gordon representation of the system, in the sense that it also admits a change of representation that casts the Hamiltonian constraint into the same Klein-Gordon equation of sLQC \cite{ACS}. In consequence, the WDW approach and LQC share the physical Hilbert space when employing the solvable Klein-Gordon formulation. In this formulation the theories differ in the representation of the observables, but the domain of the volume in the WDW approach provides that of sLQC. The nice property of the WDW approach is that it is also easily solvable in the $v$-representation (for which the question of the domain of the volume is easy to answer). We exploit this fact to arrive to the desired result: being able to write explicitly physical states in the domain of the volume in the Klein-Gordon formulation in terms of physical profiles in the $v$-representation, both for WDW and LQC. We remark that, even though the physical Hilbert space of WDW and LQC in $v$-representation are different, this poses no obstruction for our analysis, as we will show.

The structure of this work is as follows. In section \ref{sec:Hamiltonian form} we review the Hamiltonian formulation of the model suited to the quantization procedure of LQC. In section \ref{sec:WDW}, we apply the WDW quantization, analyzing its $v$-representation in \ref{sec:WDW kinematics v}, its Klein-Gordon formulation in \ref{sec:KG formulation WDW} and finding the physical relation between the two representations in \ref{subsec:Relation between forms}. In section \ref{sec:LQC} we review the quantization of the system in the context of LQC, briefly reviewing the $v$-representation and then focusing on the solvable formulation, where we write explicit physical states of the domain of the volume, with the knowledge gathered in the previous section. In section \ref{sec:semiclass}, we focus our attention on semiclassical states, clarifying the notion of semiclassicality in the solvable formulation of LQC, and studying in particular Gaussian profiles, demonstrating that now these states can be computed explicitly in this representation. Finally, in section \ref{sec:discussion} we conclude with a discussion of our work. 

We choose units $c=\hbar=1$.

\section{Hamiltonian formulation}\label{sec:Hamiltonian form}

In this work, we restrict the discussion to homogeneous and isotropic cosmological models. We start by briefly reviewing the Hamiltonian formulation of such systems suited to the quantization procedure of LQC, which mimics the techniques of LQG. For further details we refer the reader to e.g. the reviews \cite{LQCreview_Bojowald2005, LQCreview_Banerjee2012, LQCreview_Agullo2016, LQCreview_Ashtekar2011, Ashtekar_Bojowald_Lewandowski}.

Following LQG, we describe the gravitational sector of the system with a SU(2) connection, and its canonically conjugate densitized triad . Given homogeneity and isotropy, in a flat FLRW model, these variables can be parameterized by one spatially constant variable each. To this end, the connection is written as proportional to a variable $c$ and the densitized triad to a variable $p$, such that $c$ and $p$ form a canonically conjugate pair, with Poisson brackets $\lbrace c,p \rbrace = 8 \pi G \gamma/3$, where $\gamma$ is a parameter of the quantization called the Immirzi parameter \cite{LQCreview_Ashtekar2011,Immirzi}. The variable $p$ is related to the usual scale factor $a(t)$ via $a(t)=\sqrt{|p(t)|}V_o^{-1/3}$. Here we are introducing a compact cell $\mathcal{V}$ of the Universe with volume $V_o$ measured using a fiducial Euclidean metric. 
Then, the sign of $p$ depends on the relative orientation of the triad with respect to the fiducial Euclidean triad. On the other hand, $c$ is proportional to the Hubble parameter, and such that the above Poisson bracket, and subsequent physical results, do not depend on the choice of fiducial cell $\mathcal{V}$.

As in LQG, in LQC there is no operator directly representing the connection, and one represents instead its holonomies, along with fluxes of the densitized triad. The holonomies are taken along straight lines and in the fundamental representation of SU(2), with a length such that the square formed by them has an area equal to $\Delta$, the non-vanishing minimum allowed by LQG, following the \textit{improved dynamics} procedure \cite{APS_improvedDyn,APS_improvedDyn_PRD}. The fluxes, on the other hand, are simply proportional to $p$.

To simplify calculations, it is common to make a change of variables to a new canonical set, such that the holonomies simply produce a constant shift in the new geometrical variable $v$, that replaces $p$:
\begin{align}
	v &= \text{sign}(p) \frac{|p|^{3/2}}{2 \pi G \gamma \sqrt{\Delta}},\\
	b &= \sqrt{\frac{\Delta}{|p|}} c,
\end{align}
having $\lbrace b,v \rbrace=2$. This way, the volume of the cell $\mathcal{V}$ in this spacetime is given by $V = 2\pi G \gamma \sqrt{\Delta}|v|$.

Additionally, the matter content is described by a massless scalar field $\phi$ and its conjugate momentum $\pi_{\phi}$, which form a canonically conjugate pair with Poisson brackets $\lbrace \phi, \pi_{\phi} \rbrace = 1$.

Classically, the Hamiltonian of GR is found to be a linear combination of constraints. In this cosmological model, due to homogeneity, only a global Hamiltonian constraint survives:
\begin{equation}\label{eq:classical_Ham_constraint}
	\pi_{\phi}^2 - \frac{3}{4 \pi G \gamma^2} \Omega_0^2 = 0,
\end{equation}
where $\Omega_0 = 2 \pi G \gamma b v$. 

The next step is to promote the constraint \eqref{eq:classical_Ham_constraint} to a well-defined operator on a kinematical Hilbert space, by promoting the phase space variables to operators that provide representations of the canonical commutation relations. The kinematical Hilbert space will be given by two sectors: one for geometry and another one for matter.
The quantum representation of the geometrical sector is where LQC distinguishes itself from other quantum cosmology procedures, such as the WDW approach.

\section{Wheeler-De Witt approach}\label{sec:WDW}

In this section we analyse in detail the physical Hilbert space and physical states in the WDW theory and track its transformation through the different representations that lead to the Klein-Gordon formulation of the system. This will allow us to explicitly construct physical states (and in particular semiclassical ones) in the Klein-Gordon representation, both in the WDW approach and later in LQC.

\subsection{$v$-representation}\label{sec:WDW kinematics v}

Let us start by reviewing the work of \cite{APS_improvedDyn}, establishing the kinematical and physical Hilbert spaces in the representation where $\hat{v}$ is diagonal. In this setting, one adopts a standard Schr\"odinger-like representation, namely, operators $\hat{v}$, $\hat{\phi}$ act by multiplication, while $\hat{b}=2i\partial_v$ and $\hat{\pi}_{\phi}=-i\partial_\phi$, so that we obtain a representation of the canonical commutation relations: $[\hat{b},\hat{v}] = 2i$, $[\hat{\phi},\hat{\pi}_{\phi}]= i$. The quantum counterpart of the Hamiltonian constraint \eqref{eq:classical_Ham_constraint} is then given by
\begin{equation}\label{eq:constraintWDW}
	\underline{\hat{C}}=-\partial^2_{\phi} - \underline{\hat{\Theta}},
\end{equation}
where we use underlines when referring to operators/ states of the WDW approach, that are not the same as in the LQC approach. The geometric part of the constraint is
\begin{align}
	\underline{\hat{\Theta}} &= \frac{3}{4\pi G}\, \underline{\hat{\Omega}}_0^2,\\
	\underline{\hat{\Omega}}_0 &= -i\,4\pi G\ \colon v \partial_v \colon\,
\end{align}
where $\colon v \partial_v \colon$ represents the symmetric ordering of $v \partial_v$. We will use here the convenient symmetric ordering
\begin{equation}
	\colon v \partial_v \colon = \frac{v\partial_v+\partial_v v}{2},
\end{equation}
which leads to:
\begin{equation}
	\underline{\hat{\Omega}}_0 = -i\,4\pi G\left( v\partial_v + \frac{1}{2} \right).
\end{equation}

The operator $\underline{\hat{\Omega}}_0$ is (essentially) self-adjoint in the domain $\mathcal{D}_v$ of the Schwartz space of rapidly decreasing functions dense in $L^2(\mathds{R},dv)$, with absolutely continuous and non-degenerate spectrum $\sigma(\underline{\hat{\Omega}}_0)=\mathds{R}$. Its generalized eigenfunctions $\underline{e}_k(v)$ with eigenvalue $\omega_0 = 4\pi G k$ are:
\begin{equation}
	\underline{e}_k(v) = \frac{1}{\sqrt{2\pi |v|}} e^{i k \ln|v|},
\end{equation}
normalized such that $\langle \underline{e}_k,\underline{e}_{k'} \rangle = \delta(k-k')$, where $\langle\ ,\ \rangle$ is the inner product in $L^2(\mathds{R},dv)$, and $\delta(k-k')$ is the Dirac delta. This way, these eigenfunctions provide a basis for $L^2(\mathds{R},dv)$.

The operator $\underline{\hat{\Theta}}$ is (essentially) self-adjoint in $\mathcal{D}_v$, with absolutely continuous and double degenerate spectrum $\sigma(\underline{\hat{\Theta}})=\mathds{R}^+$. For each eigenvalue $\omega^2 = 12\pi G k^2$, there are two eigenfucntions: $\underline{e}_k(v)$ and its complex conjugate $\underline{e}^*_k(v)$.

Finally, the operator $-\partial_{\phi}^2$ is (essentially) self-adjoint in the domain $\mathcal{D}_{\phi}$ of the Schwartz space of rapidly decreasing functions dense in $L^2(\mathds{R},d\phi)$, with absolutely continuous double degenerate spectrum $\sigma(-\partial_{\phi}^2)=\mathds{R}^+$. Its generalized eigenfunctions of eigenvalue $\lambda^2$ are the plane waves $e^{\pm i\lambda \phi}$, which then provide a basis for $L^2(\mathds{R},d\phi)$.

Thus, the constraint operator $\underline{\hat{C}}$ is defined in a dense domain $\mathcal{D} \equiv \mathcal{D}_v \otimes \mathcal{D}_{\phi} \subset L^2(\mathds{R},dv) \otimes L^2(\mathds{R},d\phi) \equiv \mathcal{H}_{kin}$, where $\mathcal{H}_{kin}$ is the kinematical Hilbert space. Each of the operators act as the identity in the sector where they do not have a dependence.

\subsubsection{Physical states}

Physical states $\underline{\psi}$ are the ones that are annihilated by the constraint: $\underline{\hat{C}} \underline{\psi} = 0$. Since $\underline{\hat{C}}$ has \textit{continuous} spectrum, non-trivial solutions $\underline{\psi}$ are not normalizable in $\mathcal{H}_{kin}$, and we need to regard them as elements of a bigger space, specifically the topological dual $\mathcal{D}^*$ of the domain. Imposing the constraint \eqref{eq:constraintWDW} in $\mathcal{D}^*$, one finds that the physical states are generally given by:
\begin{align}\label{eq:physical states}
	\underline{\psi}(v,\phi) = \int_{-\infty}^{+\infty} dk \Big[& \tilde{\underline{\psi}}_+(k)\underline{e}_k(v) e^{i\omega(k)\phi} \nonumber\\ &+ \tilde{\underline{\psi}}_-(k)\underline{e}^*_k(v) e^{-i\omega(k)\phi} \Big],
\end{align}
where $\omega(k) = \sqrt{12\pi G} |k| > 0$.

Furthermore, interpreting $\phi$ as an internal time, physical states can be separated into positive and negative frequency sectors:
\begin{align}
	\label{eq:profiles_positiveFreq}
	\underline{\psi}_+(v,\phi) &= \int_{-\infty}^{+\infty} dk \tilde{\underline{\psi}}_+(k) \underline{e}_k(v) e^{i\omega(k)\phi},\\
	\label{eq:profiles_negativeFreq}
	\underline{\psi}_-(v,\phi) &= \int_{-\infty}^{+\infty} dk \tilde{\underline{\psi}}_-(k) \underline{e}^*_k(v) e^{-i\omega(k)\phi},
\end{align}
respectively.

Deparametrizing the system with respect to the time $\phi$, we can write a Schr\"{o}dinger evolution equation driven by the Hamiltonian $\mathH=\sqrt{\underline{\hat{\Theta}}}$ for the positive-frequency sector and by $\mathH=-\sqrt{\underline{\hat{\Theta}}}$ for the negative-frequency sector. Thus, the solutions of the model can be found by evolving an initial datum (at initial `time' $\phi = \phi_o$) with the evolution generated by $\mathH$:
\begin{equation}\label{eq:evolution_inititaldatum}
	\underline{\psi}_{\pm}(v,\phi) = e^{\pm i \sqrt{\underline{\hat{\Theta}}}(\phi-\phi_o)} \underline{\psi}_{\pm}(v,\phi_o).
\end{equation}

Before completing the physical picture in $v$-representation by providing a physical inner product and a complete set of (essentially) self-adjoint observables, let us make the observation that there is a large gauge symmetry present in the system \cite{APS_improvedDyn}: the dynamics is invariant under the change of triad orientation, and we can restrict the study to the subspace of solutions \eqref{eq:physical states} that are symmetric under the change $v\rightarrow -v$. Physical observables will preserve this space.

\subsubsection{Physical observables and physical inner-product}

Finally, we endow the space of physical states with a Hilbert space structure, by finding a complete set of commuting observables along with a physical inner product, i.e., an inner product that makes them (essentially) self-adjoint. This complete set of commuting observables consists of the constant of motion $\hat{\pi}_{\phi}$ and the relational observable $|\hat{v}|_{\phi}$ \cite{APS_improvedDyn}, where the action of $|\hat{v}|_{\phi}$ is found by separating the physical states in positive and negative frequency sectors, acting on an initial datum (at $\phi = \phi_o$) with $|\hat{v}|$, and evolving it through (\ref{eq:evolution_inititaldatum}):
\begin{equation}
\begin{aligned}
	|\hat{v}|_{\phi} \underline{\psi}(v,\phi) = &\  e^{i \sqrt{\underline{\hat{\Theta}}}(\phi-\phi_o)} |v| \underline{\psi}_+(v,\phi_o) \\
	&+ e^{-i \sqrt{\underline{\hat{\Theta}}}(\phi-\phi_o)} |v| \underline{\psi}_-(v,\phi_o).
\end{aligned}
\end{equation}

Since both these operators preserve the positive and negative frequency sectors, they are superselected, and we can restrict our analysis to one of them. We will choose to focus on the positive-frequency sector.

The physical inner-product that makes these operators (essentially) self-adjoint is:
\begin{equation}\label{dv}
	(\underline{\psi}_1,\underline{\psi}_2) = \int_{\phi=\phi_o} dv\ \underline{\psi}^*_1(v,\phi_o) \underline{\psi}_2(v,\phi_o),
\end{equation}
which is independent of the value of $\phi$, and so we evaluate it at, e.g., $\phi_o$. Recalling the invariance under triad orientation reversal, we conclude that the physical Hilbert space is $L^2_S(\mathds{R},dv)$, the symmetric part of $L^2(\mathds{R},dv)$. This way, the norm of a positive-frequency physical state is
\begin{equation}
	||\underline{\psi}_+||^2 = \int_{-\infty}^{+\infty} dv\, |\underline{\psi}_+(v,\phi)|^2= \int_{-\infty}^{+\infty} dk\, |\underline{\tilde{\psi}}_+(k)|^2.
\end{equation}
We thus arrive at the further conclusion that physical states of positive-frequency are fully characterized by the physical profiles $\underline{\tilde{\psi}}_+(k)\in  L^2(\mathds{R},dk)$. 

Analogously, the negative-frequency sector is given by profiles $\underline{\tilde{\psi}}_-(k)\in  L^2(\mathds{R},dk)$.

Moreover, in this $v$-representation the domain of the volume is defined by states $\underline\psi(v,\phi)$ specified by profiles $\underline{\tilde\psi}(k)$ that belong to $\mathcal{D}_{k}$, the Schwartz space of rapidly decreasing functions dense in $L^2(\mathds{R},dk)$.

At this stage, we anticipate that in the case of LQC, even though \eqref{dv} does not provide the physical inner product in $v$-representation, we will also find that physical states (symmetric under triad orientation reversal) are characterized by positive and negative frequency profiles ${\tilde{\psi}}_\pm(k)\in  L^2(\mathds{R},dk)$.

\subsection{Klein-Gordon formulation}\label{sec:KG formulation WDW}

Now, we can review the procedure of \cite{ACS} in detail, relating the $v$-representation with the Klein-Gordon one at the kinematical level. To this end, some intermediate changes of representation are required.

\subsubsection{$b$-representation}

Through a Fourier transformation, we pass from the $v$-representation to the conjugate $b$-representation, mapping $\underline{\psi}(v,\phi) \in L^2_S(\mathds{R},dv)$ to $\underline{\tilde{\psi}}(b,\phi) \in L^2_S(\mathds{R},db)$:
\begin{align}
	\label{eq:v to b}
	\underline{\tilde{\psi}}(b,\phi) &= \frac{1}{\sqrt{4\pi}} \int_{-\infty}^{+\infty} dv\, e^{\frac{i}{2}vb}\,\underline{\psi}(v,\phi),\\
	\underline{\psi}(v,\phi) &= \frac{1}{\sqrt{4\pi}} \int_{-\infty}^{+\infty} db\, e^{-\frac{i}{2}vb}\,\underline{\tilde{\psi}}(b,\phi).\label{eq:b to v}
\end{align}

This way, $v\partial_v$ transforms to $(-1-b\partial_b)$ and the operator $\underline{\hat{\Omega}}_0$ becomes:
\begin{equation}
	\underline{\hat{\Omega}_0} = i\,4\pi G\left(b\partial_b+\frac{1}{2}\right),
\end{equation}
defined in the Fourier transform of $\mathcal{D}_v$, namely $\tilde{\mathcal{D}}_v \subset L^2_S(\mathds{R},db)$.

For later convenience, we choose to work with the equivalent Hilbert space $L^2(\mathds{R}^+,db)\cong L^2_S(\mathds{R},db/2)$, so that we restrict the study to the positive $b$-half line in $b$-representation. Equivalently, in $v$-representation we restrict the analysis to the positive $v$-half line and from now onwards we consider physical states $\underline{\psi}(v,\phi)\in L^2(\mathds{R}^+,dv)\cong L^2_S(\mathds{R},dv/2)$, but still characterized via \eqref{eq:profiles_positiveFreq}-\eqref{eq:profiles_negativeFreq} by profiles $\underline{\tilde{\psi}}_\pm(k)$ normalized in $L^2(\mathds{R},dk)$.

\subsubsection{Rescaling}

Keeping in mind that our goal is to change variable to one that transforms the constraint into a Klein-Gordon equation, we will now perform another change of representation, in order to obtain $\underline{\hat{\Theta}} = -\sqrt{12\pi G} \left(b \partial_b \right)^2$. This can be accomplished by rescaling the states $\underline{\tilde{\psi}}(b,\phi)\in L^2(\mathds{R}^+,db)$:
\begin{equation}\label{eq:rescaling1}
	\underline{\tilde{\psi}}(b,\phi) =\frac{1}{(12\pi G)^{1/4}} \frac{1}{\sqrt{b}} \underline{\tilde{\chi}}(b,\phi),
\end{equation}
with $\underline{\tilde{\chi}}(b,\phi) \in L^2(\mathds{R}^+,\frac{1}{\sqrt{12\pi G}b}db)$. The constant factor in the above rescaling is for later convenience. This way, the constraint now reads
\begin{equation}
	\partial_{\phi}^2 \underline{\tilde{\chi}}(b,\phi) = 12\pi G (b\partial_b)^2 \underline{\tilde{\chi}}(b,\phi),
\end{equation}
and the physical inner product is
\begin{equation}
\begin{aligned}
	( \underline{\psi}_1, \underline{\psi}_2 ) &= \int_{0}^{+\infty} db\ \underline{\tilde{\psi}}^*_1(b,\phi) \underline{\tilde{\psi}}_2(b,\phi) \\
	&= \int_{0}^{+\infty} \frac{1}{\sqrt{12\pi G}} \frac{db}{b}\ \underline{\tilde{\chi}}^*_1(b,\phi) \underline{\tilde{\chi}}_2(b,\phi)=( \underline{\tilde\chi}_1, \underline{\tilde\chi}_2 ).
\end{aligned}
\end{equation}

\subsubsection{$y$-representation}

Now, we can make another change of representation, by changing from $b$ to the related variable $y$ \cite{ACS}:
\begin{align}
	y &= \frac{1}{\sqrt{12 \pi G}} \ln \frac{b}{b_o}, \\
	\label{eq:b to y}
	b &= b_o e^{\sqrt{12 \pi G} y},
\end{align}
with $b_o$ a positive constant. In the $b$-representation, this constant plays no role and physical results cannot depend on it. In fact, different choices for the value of $b_o$ correspond to unitarily equivalent theories. For convenience, we choose $b_o = 2$.

The inner product in this representation reads
\begin{equation}
\label{eq:inner product first y states}
	( \underline{\psi}_1, \underline{\psi}_2 )=( \underline{\tilde{\chi}}_1, \underline{\tilde{\chi}}_2 ) = \int_{-\infty}^{+\infty} \ dy\ \underline{\tilde{\chi}}^*_1(y,\phi) \underline{\tilde{\chi}}_2(y,\phi),
\end{equation}
with $\underline{\tilde{\chi}}(y,\phi)=\underline{\tilde{\chi}}(b(y),\phi)$.
We remind that the relation between this profile in $y$-representation $\underline{\tilde{\chi}}(y,\phi)\in L^2(\mathds{R},dy)$ and the original one in $v$-representation $\underline{\psi}(v,\phi) \in L^2_S(\mathds{R}^+,dv)\cong L^2_S(\mathds{R},dv/2)$ is obtained by using  \eqref{eq:b to v}, \eqref{eq:rescaling1}, and \eqref{eq:b to y}.

\subsubsection{Physical states}

With this change of variables, the constraint gets transformed into a Klein-Gordon equation \cite{ACS}:
\begin{equation}
	\partial_{\phi}^2 \underline{\chi}(y,\phi) = \partial_y^2 \underline{\chi}(y,\phi).
\end{equation}

The precise relation between these solutions $\underline{\chi}(y,\phi)$ and the states $\underline{\tilde{\chi}}(y,\phi)$ of the above section will be made clear in Sec. \ref{subsec:Relation between forms}, where we emphasize the novelty of this work.

The solution to the Klein-Gordon equation can be split into left and right moving modes $\underline{\chi}_L(y_+)$ and $\underline{\chi}_R(y_-)$, respectively, where $y_{\pm} = \phi \pm y$, and in positive and negative frequency sectors. We will denote the corresponding frequency by $\tilde\omega$. Focusing on positive-frequency solutions only \cite{ACS}:
\begin{align}
	\underline{\chi}(y,\phi) &= \underline{\chi}_L(y_+) + \underline{\chi}_R(y_-),\\
	\label{eq:phys_statesL}
	\underline{\chi}_L(y_+) &= \frac{1}{\sqrt{2\pi}} \int_0^{+\infty} d\tilde{\omega}e^{i\tilde{\omega}\,y_+}e^{-i\tilde{\omega}\phi_o}\underline{\tilde{\chi}}(-\tilde{\omega}),\\
	\underline{\chi}_R(y_-) &= \frac{1}{\sqrt{2\pi}} \int_0^{+\infty} d\tilde{\omega}e^{i\tilde{\omega}\,y_-}e^{-i\tilde{\omega}\phi_o}\underline{\tilde{\chi}}(\tilde{\omega}),
\end{align}
where the factor $e^{-i\tilde{\omega}\phi_o}$ was introduced for convenience, to match initial data with the previous formulation, such that the initial datum is $\underline{\chi}(y,\phi_o) = \frac{1}{\sqrt{2\pi}} \int_{-\infty}^{+\infty} d\tilde{\omega}e^{-i\tilde{\omega}y}\underline{\tilde{\chi}}(\tilde{\omega})$.

Note that $\underline{\chi}_{L/R}(y_{\pm})$ can be any function which Fourier transform is supported on the positive real line.

\subsubsection{Physical inner product}

The physical inner product in this representation is the Klein-Gordon product \cite{ACS}, namely:
\begin{equation}\label{eq:KGproduct}
\begin{aligned}
	( \underline{\chi}_1,\underline{\chi}_2 ) &= 2 \int_{-\infty}^{+\infty} d\tilde{\omega} |\tilde{\omega}| \underline{\tilde{\chi}}^*_1(\tilde{\omega}) \underline{\tilde{\chi}}_2(\tilde{\omega})\\
	&= 2\int_{-\infty}^{+\infty} dy\ \underline{\chi}^*_1(y,\phi_o) \left| i\partial_y \right| \underline{\chi}_2(y,\phi_o),
\end{aligned}
\end{equation}
where $i\partial_y$ is a positive-definite self-adjoint operator on right-moving modes and a negative definite self-adjoint operator on left-moving modes. Furthermore, the Klein-Gordon inner-product is independent of the value of $\phi$, and we have particularized it to, e.g., $\phi=\phi_o$.

Since the left and right moving sectors of this physical Hilbert space are mutually orthogonal, we can focus our analysis on the left-moving modes:
\begin{equation}
	( \underline{\chi}_1,\underline{\chi}_2)_{L} = -2i \int_{-\infty}^{+\infty} dy\ \underline{\chi}^*_{1_L}(y_+) \partial_y \underline{\chi}_{2_L}(y_+)\Big|_{\phi=\phi_o},
\end{equation}
keeping in mind that the analysis for the right-moving modes is analogous (with a plus sign in the inner product).

\subsubsection{Volume observable}
In $y$-representation the volume operator $\hat{v}$ is represented by the (essentially) self-adjoint part of
\begin{equation}
    -2 i\partial_b= \frac{1}{\sqrt{12\pi G}}e^{\sqrt{12\pi G} y}(-i \partial_y)
\end{equation}
On left-moving modes (\ref{eq:phys_statesL}), for which the operator $-i \partial_y$ is (essentially) self-adjoint and positive-definite, and with the Klein-Gordon product (\ref{eq:KGproduct}), the expectation value of the volume observable for all physical states is given by
\begin{equation}
	\label{eq:bigVexp y-rep}
	(\underline{\chi}, \hat{V}|_{\phi} \underline{\chi} )_{L} = 2 \pi G \gamma \sqrt{\Delta}\,v_\star\,e^{\sqrt{12\pi G}\ \phi},
\end{equation}
 where $v_\star$ is a state-dependent constant defined as
\begin{equation}
	\label{eq:vo y-rep}
	v_\star \equiv \frac{1}{\sqrt{3\pi G}}\int_{-\infty}^{+\infty} dy_+ \left| \frac{d\underline{\chi}_L(y_+)}{dy_+} \right|^2 e^{-\sqrt{12\pi G}\ y_+}.
\end{equation}
Note then that left-moving modes give rise to universes that expand as the time $\phi$ increases. Analogously, right-moving modes correspond to contracting universes.

In view of \eqref{eq:bigVexp y-rep}-\eqref{eq:vo y-rep}, we find that, for a state $\underline{\chi}_L(y_+)$ to belong to the domain of the volume, it has to be such that (\ref{eq:vo y-rep}) is well defined (i.e., the integral converges). Recall from \eqref{eq:phys_statesL} that it also needs to have Fourier transform with support on the positive real line to be well-defined as an element of the Hilbert space. Then, the choice of such a function is far from trivial. The solution that we propose to provide explicit physical states that belong to the domain of the volume in this Klein-Gordon formulation is to find the map between the $v$-representation and this one. 
More concretely, states $\underline\psi(v,\phi)$, specified by profiles $\underline{\tilde\psi}(k)\in \mathcal{D}_{k}$, define the domain of the volume in $v$-representation, so we seek for the relation between $\underline{\tilde\psi}(k)\in \mathcal{D}_{k}\subset L^2(\mathds{R},dk)$ and 
$\underline{\chi}_L(y_+)$.

\subsection{Relation between formulations}\label{subsec:Relation between forms}

In this section, we will build a clear dictionary at the physical level between the $v$ and $y$-representations, in order to determine the $\underline{\chi}_L(y_+)$ that corresponds to a given $\underline{\tilde{\psi}}(k)$. Given a positive-frequency profile $\underline{\tilde{\psi}}(k) \in L^2(\mathds{R},dk)$, the corresponding positive-frequency state  $\underline{\psi}(v,\phi) \in L^2(\mathds{R}^+,dv)$ in the $v$-representation is:
\begin{equation}\label{eq:phys_states_v}
	\underline{\psi}(v,\phi) = \int_{-\infty}^{+\infty}dk\ \underline{\tilde{\psi}}(k) \underline{e}_k(v) e^{i \omega(k)\,\phi}.
\end{equation}

Through (\ref{eq:v to b}), we change from the $v$-representation to the $b$-representation and obtain:
\begin{equation}
\begin{aligned}
	\underline{\tilde{\psi}}(b,\phi) =& \frac{1}{\pi \sqrt{b}} \int_{-\infty}^{+\infty}dk\ \underline{\tilde{\psi}}(k)\, e^{i \omega(k)\,\phi}\\
	& \times\left(\frac{2}{b}\right)^{i\,k} \cos\left(\frac{1+2ik}{4}\pi \right)\,\Gamma\left(\frac{1}{2}+ik\right),
\end{aligned}
\end{equation}
having $\underline{\tilde{\psi}}(b,\phi) \in L^2(\mathds{R}^+,db)$.

Then, we perform the rescaling \eqref{eq:rescaling1} and the change of variable from $b$ to $y$ through \eqref{eq:b to y}, to obtain the states $\underline{\tilde{\chi}}(y,\phi) \in L^2(\mathds{R}, dy)$. 
These states can be also split into right and left moving modes $\underline{\tilde{\chi}}(y,\phi) = \underline{\tilde{\chi}}_L(y_+) + \underline{\tilde{\chi}}_R(y_-)$:
\begin{equation}
\begin{aligned}
	\underline{\tilde{\chi}}_{L/R}(y_{\pm}) =& \frac{(12\pi G)^{1/4}}{\pi}\int_0^{+\infty} dk\ \underline{\tilde{\psi}}(\mp k)\, e^{i \omega(k)\,y_{\pm}} \\
	& \times \cos\left(\frac{1\mp 2ik}{4}\pi \right)\ \Gamma\left(\frac{1}{2}\mp ik\right) .
\end{aligned}
\end{equation}

As pointed out before, these still are not the states $\underline{\chi}(y,\phi)$. Remember that $\underline{\chi}(y,\phi)$ are not normalizable in $L^2(\mathds{R}, dy)$ but in the space with the Klein-Gordon product (\ref{eq:KGproduct}). 

Focusing e.g. on left-moving modes, in order to relate the state  $\underline{\tilde{\chi}}_L(y_+)$, which is characterized by the profile $\underline{\tilde{\psi}}(-k)\in L^2(\mathds{R},dk)$, with $\underline{{\chi}}_L(y_+)$, we take another Fourier transform:
\begin{equation}
	\underline{\tilde{\chi}}_L(\tilde{\omega},\phi) = \frac{1}{\sqrt{2\pi}} \int_{-\infty}^{+\infty} dy e^{-i\tilde{\omega}y} \underline{\tilde{\chi}}_L(y_+) \in L^2(\mathds{R},d\tilde{\omega}),
\end{equation}
and another rescaling
\begin{equation}
	\underline{\chi}_L(\tilde{\omega},\phi) = \frac{1}{\sqrt{2|\tilde{\omega}|}} \underline{\tilde{\chi}}_L(\tilde{\omega},\phi) \in L^2(\mathds{R},2|\tilde{\omega}| d\tilde{\omega}),
\end{equation}
which is finally Fourier-transformed back into
\begin{equation}
	\underline{\chi}_L(y_+) = \frac{1}{\sqrt{2\pi}} \int_{-\infty}^{+\infty} d\tilde{\omega} e^{i\tilde{\omega}y} \underline{\chi}_L(\tilde{\omega},\phi).
\end{equation}

Putting everything together, we find that the left-moving modes of the physical states of the Klein-Gordon representation are given by:

\begin{equation}
\begin{aligned}
	\underline{\chi}_L(y_+) =& \frac{1}{\sqrt{2}\pi} \int_0^{+\infty} \frac{dk}{\sqrt{k}}\ \underline{\tilde{\psi}}(-k) \cos\left(\frac{1-2ik}{4}\pi \right)\\
	& \times\Gamma\left(\frac{1}{2}-ik\right)e^{i\sqrt{12\pi G}\,k\,y_+}.
\end{aligned}
\end{equation}

\subsection{Summary}

In summary, in the Klein-Gordon representation, the physical states $\underline{\chi}(y,\phi)$ can be split in left and right moving modes:
\begin{equation}\label{eq:L and R modes of physical state in KG}
	\underline{\chi}_{L/R}(y_{\pm}) = \frac{1}{\sqrt{2\pi}} \int_0^{+\infty} dk\,\underline{\tilde{\chi}}_{\pm}(k)\,e^{i\sqrt{12\pi G}\,k\,y_{\pm}},
\end{equation}
respectively, whose Fourier transform $\underline{\tilde{\chi}}_{\pm}(k)$ have support on the positive real line. These Fourier transforms are the profiles that define the physical state, and are related to the profile $\underline{\tilde{\psi}}(k) \in L^2(\mathds{R},dk)$ defining \eqref{eq:phys_states_v} by:

\begin{equation}
	\label{eq:Fourier transform relation to v profile}
	\underline{\tilde{\chi}}_{\pm}(k) = \frac{1}{\sqrt{\pi}} \frac{1}{\sqrt{k}}\ \underline{\tilde{\psi}}(\mp k) \cos\left(\frac{1\mp 2ik}{4}\pi \right)\ \Gamma\left(\frac{1}{2}\mp ik\right).
\end{equation}

Thus, the connection between the $v$-representation and the Klein-Gordon one is found at the physical level. In particular, this allows the computation of the expectation value of the volume from (\ref{eq:bigVexp y-rep}), given a profile in its dense domain $\underline{\tilde{\psi}}(k)\in \mathcal{D}_{k}\subset L^2(\mathds{R},dk)$.

\section{Loop Quantum Cosmology}
\label{sec:LQC}

\subsection{$v$-representation}\

Following the techniques of LQG, there is no operator representing the connection directly, but instead the holonomies of the connection are represented with the operator $\widehat{e^{i b/2}}$. In $v$-representation, this operator produces a constant shift in the variable $v$, providing a representation of the canonical commutation relation $[\hat{v},\widehat{e^{i b/2}}]=\widehat{e^{i b/2}}$. Remarkably, the inner product is discrete, given by the Kronecker delta (instead of the Dirac delta, as in the WDW approach):
\begin{equation}\label{eq:Hgrav_innerprod}
	\langle v | v' \rangle = \delta_{v,v'},
\end{equation}
where the quantum states are represented as $|v\rangle$. This is a consequence of the fact that in this representation there is no infinitesimal generator $\hat{b}$ of translations in $v$, but only finite translations $\widehat{e^{i b/2}}$ are well defined. Thus, the basis states $|v\rangle$ are normalizable and provide an orthonormal basis.

Explicitly, the quantum counterpart of the Hamiltonian constraint is given by:
\begin{equation}
	\hat{C}=-\partial^2_{\phi} - \hat{\Theta}.
\end{equation}
The geometric part of the constraint is given by $\hat{\Theta} = \frac{3}{4 \pi G \gamma^2} \hat{\Omega}_0^2$, and $\hat{\Omega}_0$ is the symmetric operator \cite{Merce_Guillermo_Javier}:
\begin{equation}
	\hat{\Omega}_0 = \frac{1}{2\sqrt{\Delta}} \hat{V}^{1/2} \left[ \widehat{\text{sign}(v)}\ \widehat{\sin b} + \widehat{\sin b}\ \widehat{\text{sign}(v)} \right]\hat{V}^{1/2}.
\end{equation}
where $\widehat{\sin b} = \left( \widehat{e^{ib}}-\widehat{e^{-ib}}\right)/(2i)$. Then, the operator $\hat{\Theta}$ is a difference operator of step 4, densely defined in the semilattices $\mathcal{L}_{\varepsilon}^{\pm}$:
\begin{equation}
	\mathcal{L}_{\varepsilon}^{\pm}= \lbrace |v\rangle = |\pm \left(\varepsilon + 4n\right)\rangle,\ n \in \mathds{N} \rbrace,\qquad \varepsilon \in (0,4],
\end{equation}
such that it is essentially self-adjoint in the Hilbert spaces $\mathcal{H}_{\varepsilon}^{\pm}$, the closure of $\mathcal{L}_{\varepsilon}^{\pm}$ with respect to the inner product \eqref{eq:Hgrav_innerprod}.

The generalized eigenfunctions of $\hat{\Theta}$, $e_k(v)$, verify a recurrence relation and do not admit a simple closed form. In the limit of large $v$, they tend to a real linear combination of the two corresponding eigenfunctions of the WDW approach for the same eigenvalue. In other words, they behave like standing waves with both outgoing and incoming components, which then do not decouple unlike in the WDW approach \cite{Merce_Guillermo_Javier}.

On the other hand, in LQC the matter field is quantized with a standard Schr\"{o}dinger-like representation, exactly as in the WDW approach.
The total kinematical Hilbert space is then $\mathcal{H}_{kin} = \mathcal{H}_{\varepsilon}^{\pm} \otimes L^2(\mathds{R},d\phi)$. Hence, as in the WDW approach, each of the operators act as the identity in the sector where they do not have a dependence.

With an analysis analogous to that exposed in section \ref{sec:WDW kinematics v} for the WDW approach, we find that the physical states are given by:
\begin{equation}
	\psi(v,\phi) = \int_{-\infty}^{+\infty} dk\ e_k(v) \left[ \tilde{\psi}_+(k) e^{i\omega(k)\phi} + \tilde{\psi}_-(k) e^{-i\omega(k)\phi} \right].
\end{equation}

Here, as in the WDW approach, the physical Hilbert space is $\mathcal{H}_{phys} = L^2(\mathds{R}, dk) \ni \tilde{\psi}_{\pm}(k)$. Thus, $\tilde{\psi}_{\pm}(k)$ provide again superselected positive and negative frequency sectors, in the same way as in the WDW approach, and again $\tilde{\psi}_{\pm}(k)\in \mathcal{D}_{k}$ provide the domain of the volume observable.

\subsection{Klein-Gordon formulation -- sLQC}

For the case of a flat FLRW model with a massless scalar field, there is a specially useful representation to work with, which allows for the constraint to be solved analytically. This is obtained by following the solvable LQC (sLQC) prescription \cite{ACS}. The procedure is similar to that exposed in section \ref{sec:KG formulation WDW}. Firstly, we change from the $v$-representation to the $b$-representation, this time, performing a \textit{discrete} Fourier transform. For this purpose, we need to define $\hat{\Theta}$ in a lattice supported over the whole real line, symmetrically spread around $v=0$:
\begin{equation}
	\mathcal{L}= \lbrace |v\rangle = |4n\rangle,\ n \in \mathds{Z} \rbrace,
\end{equation}
so that $\mathcal{L} = \left( \mathcal{L}_4^+ \cup \mathcal{L}_4^- \cup |0\rangle \right)$ (we define the states $\psi(v,\phi)$ to vanish at $v=0$). This way, the Hilbert space under consideration for the geometry sector, $\mathcal{H}_{grav}$, is the closure of $\mathcal{L}$ with respect to the inner-product \eqref{eq:Hgrav_innerprod}. As in the WDW approach, the reversal of the triad orientation, $v\rightarrow-v$, is a large gauge symmetry and we consider symmetric states. The operator $\hat{\Theta}$ is essentially self-adjoint in $\mathcal{H}_{grav}$, and remarkably its spectrum is non-degenerate, unlike its analog $\underline{\hat{\Theta}}$ in the WDW theory. This difference is the fundamental reason why the dynamics of LQC displays a bounce, instead of the two types of solutions (expanding and contracting) of the WDW approach \cite{Merce_Guillermo_Javier}.

Then, the wave functions $\psi(v,\phi)$ and $\tilde{\psi}(b,\phi)$ of the $v$ and $b$-representations are related by:

\begin{align}
	\tilde{\psi}(b,\phi) &= \sum_{v \in \mathcal{L}} e^{\frac{ibv}{2}} \psi(v,\phi),\\
	\psi(v,\phi) &= \frac{1}{\pi} \int_0^{\pi} db\ e^{-\frac{ibv}{2}} \tilde{\psi}(b,\phi).
\end{align}

Notice that this Fourier transform maps $\mathcal{H}_{grav}$ to $L^2([0,\pi],db)$ (with periodic boundary conditions), i.e., since $v$ is supported on a lattice of equidistant points over the real line, $b$ is now an angle.

Then, we introduce a scaling \cite{ACS}
\begin{equation}
    \chi(v,\phi) =\frac{\pi}{v} \psi(v,\phi) 
\end{equation}
 and the change of variable:
\begin{equation}
	x=\frac{1}{\sqrt{12\pi G}}\ln\left[ \tan\left( \frac{b}{2} \right) \right],
\end{equation}
analogous to the change of variable from $b$ to $y$ in the WDW approach. This way, the constraint gets transformed into the same Klein-Gordon equation as in the WDW approach, where now $x$ plays the role of $y$,
\begin{equation}
	\partial_{\phi}^2 {\chi}(x,\phi) = \partial_x^2 {\chi}(x,\phi).
\end{equation}
Therefore, again positive and negative frequency sectors decouple and we can restrict the study to e.g. positive-frequency states, for which $-i\partial_\phi$ is positive-definite.

Nevertheless now there is an important difference with respect to the WDW theory, and it finds its origin again in the non-degeneracy of the spectrum of $\hat\Theta$ in LQC.  In fact, now the invariance under reversal of the triad leads to $\chi(-x,\phi)=-\chi(x,\phi)$. As a consequence, physical states have the form 
\begin{equation}\label{eq:sLQC_states}
	\chi(x,\phi) = \frac{1}{\sqrt{2}}\left[\chi(x_+)-\chi(x_-)\right],
\end{equation}
where $\chi(x_{\pm}) = \chi(\phi \pm x)$ correspond to the left and right moving modes, respectively, and $\chi$ is any function with Fourier transform supported on the positive real line (to be positive-frequency). In other words, while in the WDW approach the left and right moving sectors are completely independent, in LQC they are not: left-moving modes determine right-moving modes and vice versa. In the following we specify the physical state using e.g. only left-moving modes.

The total inner product on physical states is
\begin{align}\label{eq:expvalue=sum of L and R/2}
	( \chi_1,\chi_2 )&= \frac{1}{2} \left[ ( \chi_1,\chi_2 )_{L} - ( \chi_1,\chi_2 )_{R} \right]=( \chi_1,\chi_2 )_{L}\nonumber\\
	&=-2 i \int_{-\infty}^\infty dx\ {\chi_1}^ *(x_+)\partial_x \chi_2(x_+),
\end{align}
which is independent on the time $\phi$ and we can evaluate it for example at $\phi=\phi_o$.

\subsection{Volume observable}

Now we can compute the volume observable, as we did for the WDW theory. In the present $x$-representation of LQC, the operator $\hat{v}$ is represented by the (essentially) self-adjoint part of 
\begin{equation}
    -2 i\partial_b= \frac{2}{\sqrt{12\pi G}}\cosh(\sqrt{12\pi G} x)(-i \partial_x)
\end{equation}
Let us recall that on left-moving modes $\chi(x_+)$ the operator $-i \partial_x$ is (essentially) self-adjoint and positive-definite. Then it is easy to see that the expectation value of the volume observable for all physical states, is now given by
\begin{equation}
	\label{eq:bigVexp x-rep}
	({\chi}, \hat{V}|_{\phi} {\chi} ) = 2 \pi G \gamma \sqrt{\Delta}\left(v_+e^{\sqrt{12\pi G}\ \phi}+v_-e^{-\sqrt{12\pi G}\ \phi}\right),
\end{equation}
 where $v_\pm$ are state-dependent constants defined, in terms of the left-moving modes $\chi(x_+)$, as
\begin{equation}
	\label{eq:V+- in sLQC}
	v_\pm \equiv \frac{1}{\sqrt{3\pi G}}\int_{-\infty}^{+\infty} dx_+ \left| \frac{d{\chi}(x_+)}{dx_+} \right|^2 e^{\mp\sqrt{12\pi G}\ x_+}.
\end{equation}

Equation \eqref{eq:bigVexp x-rep} makes obvious that now the presence of both components, left and right movers, leads to the occurrence of a bounce. Indeed, we can rewrite \cite{ACS}
\begin{equation}\label{eq:<V> as cosh}
	({\chi}, \hat{V}|_{\phi} {\chi} ) = 2 \pi G \gamma \sqrt{\Delta}v_B\cosh[\sqrt{12\pi G}(\phi-\phi_B)],
\end{equation}
with 
\begin{equation}
    v_B=2\frac{\sqrt{v_+v_-}}{||\chi||}, \quad \phi_B=\frac{1}{2\sqrt{12\pi G}}\ln\left(\frac{v_-}{v_+}\right),
\end{equation}
state-dependent constants giving the value of the volume and the scalar field at such bounce.

Similarly to the WDW approach, the question that we have to answer is which positive-frequency left-moving modes $\chi(x_+)$ give rise to physical states belonging to the domain of the volume observable. Namely we look for profiles $\chi(x_+)$ which Fourier transform is supported in the positive real line and such that the constants $v_\pm$ are finite. These conditions are exactly the same as in the WDW approach! 

Therefore the answer is that the domain of the volume in LQC is characterized by left-moving modes
\begin{equation}
	\label{eq:chi(x+) LQC}
	\chi(x_+) = \frac{1}{\sqrt{2\pi}} \int_0^{+\infty} dk\,\tilde{\chi}(k)\,e^{i\sqrt{12\pi G}\,k\,x_+},
\end{equation}
with Fourier transform $\tilde{\chi}(k)$ defined as
\begin{equation}
	\label{eq:tilde chi(k) LQC}
	\tilde{\chi}(k) = \frac{1}{\sqrt{\pi}} \frac{1}{\sqrt{k}}\ \tilde{\psi}(-k) \cos\left(\frac{1-2ik}{4}\pi \right)\ \Gamma\left(\frac{1}{2}-ik\right).
\end{equation}
Here $\tilde{\psi}(k) \in \mathcal{D}_{k}\subset L^2(\mathds{R},dk)$ is the physical profile that defines the positive-frequency state in $v$-representation
\begin{equation}
	\psi(v,\phi) = \int_{-\infty}^{+\infty} dk\ e_k(v)  \tilde{\psi}(k) e^{i\omega(k)\phi}.
\end{equation}

This completes the link at the physical level between the original representation of LQC and the $x$-representation of sLQC.
Note that we have managed to make the connection between both formulations without the need to solve analytically for the LQC eigenfunctions $e_k(v)$. Actually one could follow that other path, namely using the analytical solution of sLQC to solve for $e_k(v)$, as done in \cite{Ashtekar_Campiglia_Henderson,Craig}. Those works did not worry about the domain of the volume operator, that we have explicitly constructed here. Using the analytical expression for $e_k(v)$ (conveniently symmetrized and normalized) one should arrive at the same result than us. However we decided instead to exploit the parallelism with the WDW approach as we find it more convenient. 

\section{Semiclassicality}\label{sec:semiclass}

In this description, we have made no restrictions to the physical states so far. However, we are particularly interested in semiclassical states, in the interest of agreeing with classical General Relativity in the low curvature regime where quantum geometrical effects should not be important. In other words, semiclassical states are those for which the expectation values of physical observables in the low curvature regime are peaked in a classical trajectory, so that the relative dispersions of such observables are small in that regime. 

In the model under study we have two independent observables, the momentum of the scalar field, and the volume. Thus, we call localized states those for which
\begin{align}
\frac{\langle \Delta\hat{\pi}_{\phi}\rangle}{\langle \hat{\pi}_{\phi}\rangle} \,,\, \frac{\langle \Delta\hat{V}|_{\phi}\rangle }{\langle\hat{V}|_{\phi}\rangle} 
	\label{eq:semiclassicality}
\end{align}
are bounded along the evolution. The smaller these relative dispersions are the more semiclassical the states are.
Here, and in what follows, $\langle \hat{O} \rangle \equiv (\chi|\hat{O} \chi)$ denotes the expectation value of a physical observable $\hat{O}$ on the left-moving mode $\chi$.

Having the relation between the left-moving modes $\chi(x_+)$ and the spectral profiles $\tilde\psi(k)$, restricting to semiclassical states is straightforward, because we can infer from previous studies in $v$-representation which conditions $\tilde\psi(k)$ must satisfy to give rise to a semiclassical state. In general, a profile suficiently peaked on a finite value of $k$ does the job. For example, the most natural choice, which is the one investigated in \cite{APS_improvedDyn_PRD}, is a Gaussian profile
\begin{equation}
	\tilde{\psi}(k) = N e^{-\frac{(k-k_o)^2}{2\sigma^2}},
\end{equation}
centered on $k_o$, with a standard deviation $\sigma$ and a  normalization factor $N$.

In this section, we are going to forget for a moment the relation between $\chi(x_+)$ and $\tilde\psi(k)$, and analyze the question of semiclassicality directly in the solvable formulation of LQC.

For convenience, in what follows we will make the change of variable $u=\sqrt{12\pi G} x_+$, so that the left-moving modes $\chi(x_+)$ will become $\chi(u)$, and
\begin{align}\label{eq:app v pm simplified}
    v_{\pm} &=2 \int_{-\infty}^{+\infty} du \left| \frac{d\chi(u)}{du} \right|^2 e^{\mp u}.
    \end{align}

\subsection{Momentum of the field}
Focusing first on the momentum of the scalar field, in the solvable formulation of LQC one straightforwardly finds $\langle \hat{\pi}_{\phi} \rangle = \sqrt{12\pi G} v_o$, where $v_o$ is a state-dependent constant of motion defined via
\begin{equation}\label{eq:vo}
	v_o \equiv 2\int_{-\infty}^{+\infty} du \left| \frac{d\chi(u)}{du} \right|^2.
\end{equation}
To find the dispersion of the momentum of the field,
\begin{equation}
	\langle \Delta\hat{\pi}_{\phi} \rangle=\sqrt{ \langle \hat{\pi}_{\phi}^2 \rangle - \langle \hat{\pi}_{\phi} \rangle^2},
\end{equation}
we notice that another straightforward calculation leads to $\langle \hat{\pi}_{\phi}^2 \rangle = 6\pi G v_{d_0}^2$, where $v_{d_0}^2$ is another state-dependent constant in $\phi$, defined via
\begin{equation}
\begin{aligned}\label{eq:vdn}
    v^2_{d_n} = 4 \int_{-\infty}^{+\infty} du\, \text{Im}\left[ \frac{d\chi^*(u)}{du}\frac{d^2\chi(u)}{du^2}\right] e^{-nu},
\end{aligned}
\end{equation}
for the case $n=0$. 
This yields 
\begin{align}\label{eq:dis_momentum}
    \frac{\langle \Delta\hat{\pi}_{\phi}\rangle }{\langle \hat{\pi}_{\phi}\rangle}=\sqrt{\frac{v_{d_0}^2}{2v_o^2}-1}.
\end{align}

Notice that $v_o$ is very similar to $v_{\pm}$ defined in \eqref{eq:app v pm simplified}, which already have to be finite for physical states in the domain of the volume. For the integral \eqref{eq:app v pm simplified} to exist, since $e^{\mp u}$ diverges for $u\rightarrow \mp \infty$, we conclude that $|d\chi/du|^2$ is such that it dominates over $e^{\mp u}$, compensating for its divergent behavior. This implies that $|d\chi/du|^2$ converges by itself in the integration range. Therefore, $v_o$ is finite as long as the state is in the domain of the volume. In consequence,  the boundeness of the relative dispersion of the momentum of the field implies then that $v^2_{d_0}$ has to be finite as well.
Moreover, the semiclassicality condition
 ${\langle \Delta\hat{\pi}_{\phi}\rangle}/{\langle \hat{\pi}_{\phi}\rangle} \ll 1$ applies for profiles with $v_{d_0}^2/(2v_o^2) \gtrsim 1$.

\subsection{Volume}

Analogously for the volume observable, we find that the relative dispersion  can be written as
\begin{equation}
\begin{aligned}\label{eq:dispersion V}
    \frac{\langle \Delta \hat{V}|_{\phi} \rangle}{\langle \hat{V}|_{\phi} \rangle} & = \Biggl[\frac{v^2_{d_{+2}}}{2v_+^2 + 2v_-^2\, e^{-4\sqrt{12\pi G}\,\phi} + 4v_+v_-\,e^{-2\sqrt{12\pi G}\,\phi}}\\
    & + \frac{v^2_{d_{-2}}}{2v_+^2\, e^{4\sqrt{12\pi G}\,\phi} + 2v_-^2 + 4v_+v_-\,e^{2\sqrt{12\pi G}\,\phi}}\\
    & + \frac{ v^2_{d_0}}{v_+^2\,e^{2\sqrt{12\pi G}\,\phi} + v_-^2\,e^{-2\sqrt{12\pi G}\,\phi} + 2v_+v_-} - 1 \Biggr]^{1/2},
\end{aligned}
\end{equation}
where we defined $v^2_{d_\pm2}$ according to \eqref{eq:vdn}, for $n=\pm2$.

Throughout the evolution in $\phi$, the first term of \eqref{eq:dispersion V} is bounded by $v^2_{d_{+2}}/(2v_+^2)$ (which is reached when $\phi \rightarrow +\infty$), the second by $v^2_{d_{-2}}/(2v_-^2)$ (reached when $\phi \rightarrow -\infty$) and the third by $v^2_{d_0}/(2v_+v_-)$ (when $\phi =\phi_B$).
It is easy to see that finiteness of $v^2_{d_{\pm 2}}$ implies finiteness of $v^2_{d_{0}}$, applying a similar argument to that below \eqref{eq:dis_momentum}. Therefore states localized in the volume are also localized in the momentum of the field.
Moreover, the semiclassicality condition
 $ {\langle \Delta \hat{V}|_{\phi} \rangle}/{\langle \hat{V}|_{\phi} \rangle} \ll 1$ applies for profiles with $v^2_{d_{\pm2}}\gtrsim 2v_{\pm}^2$ and $v^2_{d_{0}}\gtrsim 2v_+v_-$.

\subsection{Conditions on $\chi(x_+)$ to guarantee semiclassicality}

From the above analysis, we conclude that for states $\chi(u)$ belonging to the domain of the volume, namely with $v_{\pm}<\infty$, to define states localized both in the volume and in the momentum of the field, we further need the following two conditions to hold
\begin{align}\label{eq:localization}
     v_{d_{+2}}^2<\infty\quad, \quad v_{d_{-2}}^2<\infty.
\end{align}
Moreover, semiclassical states are those highly localized in the volume observable, for which  
\begin{align}\label{eq:semiclassical vol}
    v^2_{d_{\pm2}}\gtrsim 2v_{\pm}^2\quad, \quad v^2_{d_{0}}\gtrsim 2v_+v_-,
\end{align}
as well as in the momentum of the field, for which
\begin{align}\label{eq:semiclassical momentum}
    v_{d_0}^2 \gtrsim 2v_o^2.
\end{align}

If \eqref{eq:localization} holds, we could always choose appropriately the parameters of the profile $\chi(u)$ to guarantee \eqref{eq:semiclassical vol} and \eqref{eq:semiclassical momentum}.
Let us then study what the conditions \eqref{eq:localization} impose on the behavior of $\chi(u)$ as a function of $u$.
Let us write
\begin{equation}\label{eq:dchidx with phase}
    \frac{d\chi(u)}{du} = \left| \frac{d\chi(u)}{du} \right| e^{i\alpha(u)},
\end{equation}
where $\alpha(u)$ is a real-valued function, standing for a possible phase. This way,
\begin{align}
    \label{eq:v_d_2 with alpha}
    v_{d_\pm 2}^2 &=4 \int_{-\infty}^{+\infty} du\, \alpha'(u) \left| \frac{d\chi(u)}{du} \right|^2 e^{\mp 2 u}.
\end{align}

If $v_{\pm}$ is finite, we infer that $\left| d\chi/du \right|^2$ dominates over $e^{\mp u}$ for $u \rightarrow \pm \infty$. Then, we can write
\begin{equation}
    \left| \frac{d\chi(u)}{du} \right|^2 \sim f(u)e^{-|u|},
\end{equation}
with $f(u)$ an integrable function in each semiaxis of the real line. Similarly, finiteness of $v^2_{d_{\pm 2}}$ implies the behavior 
\begin{equation}
    \alpha'(u) \sim g(u)e^{-|u|},
\end{equation}
with $g(u)$ such that $g(u)f(u)$ is integrable in each semiaxis of the real line.

\subsection{Gaussian profiles}

We will now consider the positive-frequency states analyzed in \cite{APS_improvedDyn_PRD}, that are 
characterized by a Gaussian spectral profile centered at $k_o$, with width $\sigma$:
\begin{equation}\label{eq:gaussian profile}
	\tilde{\psi}(k) = \frac{1}{\sqrt{\sigma\sqrt{\pi}}} e^{-\frac{(k-k_o)^2}{2\sigma^2}}.
\end{equation}
We will compute $|d\chi/du|^2$ and $\alpha'(u)$ for these states, as a particular example of the previous discussion.

In $v$-representation, using $\langle e_k'|e_k\rangle=\delta(k-k')$, it is straightforward to obtain 
$\langle \hat{\pi}_{\phi}\rangle=-\sqrt{12\pi G}k_o$ and
\begin{align}\label{eq:dis_momentum2}
    \left(\frac{\langle \Delta\hat{\pi}_{\phi}\rangle }{\langle \hat{\pi}_{\phi}\rangle}\right)^2=\frac{\sigma^2}{2k_o^2},
\end{align}
 on these states. We note that agreement with the solvable formulation requires $v_o=-k_o$ for e.g. the left-moving sector of these states, as we must have  $\langle \hat{\pi}_{\phi} \rangle = \sqrt{12\pi G} v_o$.
 
 We will focus our analysis on the ranges $-450 \leq k_o \leq -50$ and $0.01 \leq \sigma \leq 0.1$, so that $\sigma\ll |k_o|$. This way we have high localization in the momentum of the field. Later we will analyze the localization in the volume.

To compute $|d\chi/du|^2$, we first need to integrate
\begin{equation}
    \frac{d\chi(u)}{du} = \frac{i}{\sqrt{2\pi}} \int_0^{+\infty} dk\,k\,\tilde{\chi}(k)\,e^{i k\,u}.
\end{equation}
We recall that $\tilde{\chi}(k)$ is given in \eqref{eq:tilde chi(k) LQC} in terms of the Gaussian profile \eqref{eq:gaussian profile}. 
For each pair of parameters $(k_o,\sigma)$, we have performed  the integration numerically as we did not manage to solve the integral by analytical methods. We have taken an interval of $u$ large enough for $\left|d\chi/du\right|^2$ to converge to $0$ in the endpoints of that interval.
The integrand function is highly oscillatory, which requires sufficiently small stepsizes for the integrations.  It turns out that the final result fits to the Gaussian function
\begin{align}
   \left| \frac{d\chi(u)}{du} \right|^2 =A_o e^{-\frac{(u-p_o)^2}{2 \sigma_o^2}}.
\end{align}
We numerically find  that the parameters $A_o$, $p_o$, and $\sigma_o$ depend on the parameters $(k_o,\sigma)$ in the following way
\begin{align}
    \label{eq:app fit A0}
    {A}_o(k_o,\sigma)  & = (-0.499999971 \pm 3\cdot10^{-9}) \frac{k_o \sigma}{\sqrt{\pi}},\\
    \label{eq:app fit p0}
    p_o(k_o,\sigma) & = \ln[(-0.999987 \pm 7\cdot10^{-6}) k_o],\\
    \label{eq:app fit sigma0}
    \sigma_o(k_o,\sigma) & =\frac{0.7071066 \pm 1\cdot10^{-7}}{\sigma}.
\end{align}
Thus, our numerical analysis leads to the analytical expression
\begin{align}\label{eq:der-chi}
   \left| \frac{d\chi(u)}{du} \right|^2 =-\frac{k_o \sigma}{2\sqrt{\pi}} e^{-\sigma^2[u-\ln(-k_o)]^2},
\end{align}
valid at least in the range of parameters $-450 \leq k_o \leq -50$ and $0.01 \leq \sigma \leq 0.1$.
Moreover, replacing this result in \eqref{eq:vo} we obtain $v_o=-k_o$ as expected. This serves as a test that \eqref{eq:der-chi} is correct.

Let us now analyze $\alpha'(u)$. 
Considering \eqref{eq:dchidx with phase}, we have computed $d\chi/du$ numerically, and $\alpha(u)$ has been then obtained by
\begin{equation}\label{eq:app cos(alpha)}
  \cos[\alpha(u)] =\text{Re}\left[  \frac{d\chi}{du} / \left|\frac{d\chi}{du}\right| \right].
\end{equation}
Remarkably, we obtain that the frequency of this trigonometric function is constant along $u$.
Therefore $\alpha(u)$ is linear in $u$, resulting in constant $\alpha'$. 
Numerically, this was confirmed by applying a $\cos^{-1}$ function to the data, obtaining a sectional linear function in $u$ (actually the slope is constant but alternates sign with a given period) for the entire range of parameters studied. 
Since $\alpha'$ is constant, we find
\begin{align}
    v_{d_0}^2 &=4 \int_{-\infty}^{+\infty} du\, \alpha'(u) \left| \frac{d\chi(u)}{du} \right|^2=2\alpha'v_o.
\end{align}
On the other hand, comparing \eqref{eq:dis_momentum} with \eqref{eq:dis_momentum2} we conclude that $v_{d_0}^2=\sigma^2+2k_o^2$, so that for consistency
\begin{align}
    \alpha'=-\frac{\sigma^2+2k_o^2}{2k_o}.
\end{align}
This result is also compatible with our numerics.

Let us now analyze which further conditions have to verify $\sigma$ and $k_o$ for these states to provide localization in the volume.
We can  straightforwardly compute $v_\pm$ and $v_{d_\pm2}^2$  by relating them to $v_o$ and $v_{d_0}^2$ respectively. From \eqref{eq:app v pm simplified} and \eqref{eq:v_d_2 with alpha} we get
\begin{align}
    v_\pm=v_o e^{\mp p_o+\sigma_o^2/2}\quad,\quad v_{d_\pm2}^2=v_{d_0}^2 e^{2(\mp p_o+\sigma_o^2)}.
\end{align}
Now, using $v_o=-k_o$, $v_{d_0}^2=\sigma^2+2k_o^2$, and the fits \eqref{eq:app fit p0}-\eqref{eq:app fit sigma0},  we get $ v_+=e^{{1}/{4\sigma^2}}$, $v_-=k_o^2v_+$, and
\begin{align}
  \frac{v_{d_{+2}}^2}{2v_+^2}=\frac{v_{d_{-2}}^2}{2v_-^2}&=\left(\frac{\sigma^2}{2k_o^2}+1\right)e^{\frac{1}{2\sigma^2}},\\
  \frac{v_{d_{0}}^2}{2v_+v_-}&=\left(\frac{\sigma^2}{2k_o^2}+1\right)e^{-\frac{1}{2\sigma^2}}.
\end{align}
We conclude that semiclassicality both in the volume and in the momentum of the field requires a relatively large value of $\sigma$ while having $\sigma\ll |k_o|$.

\section{Conclusions/Discussion}\label{sec:discussion}

The dynamics of a flat FLRW spacetime minimally coupled to a massless scalar field had already been studied in the context of LQC. In the $v$-representation where the volume operator is diagonal, the dynamics was originally computed numerically \cite{APS,APS_improvedDyn,APS_improvedDyn_PRD}, which lead to the exploration of another representation. Applying the so-called \textit{solvable} prescription \cite{ACS}, the constraint is cast into a Klein-Gordon equation, where an analytical treatment is possible. However, in this solvable formulation, some details have been left unobserved in previous works. Namely, the connection between the $v$-representation and the solvable one had only been presented in \cite{ACS} at the kinematical level, and thus the domain of the volume (the main observable under consideration) was not established in this representation.

In this work, we have developed the connection at the physical level between the two representations. We have obtained this mapping first in the context of the WDW approach, where, in the equivalent $v$-representation, the dynamics is also analytically solvable. This approach admits a Klein-Gordon representation as well, which shares its physical Hilbert space with the solvable representation of LQC. Then in the WDW approach, the connection between its $v$-representation and the Klein-Gordon one is more easily found than in LQC. This way, we find an explicit form for the states in the domain of the volume in the Klein-Gordon representation of the WDW approach. These states turn out to provide as well the domain of the volume in the LQC approach.

Furthermore, we have analyzed the notion of semiclassicality in the solvable formulation of LQC. We have showed that localization in the volume implies localization in the momentum of the field and looked for profiles that defined states localized in both these observables. As in the $v$-representation, we find that a Gaussian profile defines semiclassical states, with an appropriate choice of its parameters.

Finally, let us mention that there are previous works that also looked at the connection at the physical level between solvable LQC and the original $v$-representation. In particular, \cite{Ashtekar_Campiglia_Henderson,Craig} provide the analytic expression of the eigenfunctions that diagonalize the Hamiltonian constraint operator in $v$-representation, which were only generated numerically (and employing approximations) in \cite{APS_improvedDyn,APS_improvedDyn_PRD}. More recently, \cite{Norbert} points out a disagreement between the physical inner product of both formulations. Our work makes the connection without relying on the analytic expression of the LQC eigenfunctions, but instead uses those of the WDW approach that are simpler. Therefore, our work also complements these previous ones, and might serve useful to clarify the issue pointed out in \cite{Norbert}.

\acknowledgements

M. Mart\'in-Benito thanks G. A. Mena Marug\'an and L. J. Garay 
for discussions. This work was partially
supported by the  Spanish  MINECO  grants  FIS2014-54800-C2-2-P  and
FIS2017-86497-C2-2-P.

\bibliography{LQCbib_1}



\end{document}